\documentclass[runningheads]{llncs}
\usepackage[T1]{fontenc}
\usepackage{graphicx}
\usepackage{booktabs}
\usepackage[misc]{ifsym}
\usepackage{listings}

\usepackage{mwe}
\usepackage{xcolor}

\definecolor{codegreen}{rgb}{0,0.6,0}
\definecolor{codegray}{rgb}{0.5,0.5,0.5}
\definecolor{codepurple}{rgb}{0.58,0,0.82}
\definecolor{backcolour}{rgb}{0.95,0.95,0.92}
\lstdefinestyle{mystyle}{
    backgroundcolor=\color{backcolour},   
    commentstyle=\color{codegreen},
    keywordstyle=\color{magenta},
    numberstyle=\tiny\color{codegray},
    stringstyle=\color{codepurple},
    basicstyle=\ttfamily\scriptsize,
    breakatwhitespace=false,         
    breaklines=true,                 
    captionpos=b,                    
    keepspaces=true,                 
    numbers=left,                    
    numbersep=5pt,                  
    showspaces=false,                
    showstringspaces=false,
    showtabs=false,                  
    tabsize=2
}
\begin{document}

\title{A Concept-based approach to Voice Disorder Detection}
%Voice Disorders Detection through Explainable AI. A concept-based approach}

\titlerunning{A Concept-based approach to Voice Disorder Detection}
% If the full title of your paper is short enough to also fit in the running head, you can omit the abbreviated paper title here. You can check as follows: if you comment out the \titlerunning line, something will appear in the header of all odd-numbered pages of your PDF from page 3 onward. This something is either the full title (in which case all is well), or the error message "Title Suppressed Due to Excessive Length". If this error message appears, you're going to want to provide an abbreviated title within the \titlerunning command, because if you won't do it, Springer will do it for you.

%N.B.: Author information (both in the \author{} and \authorrunning{} command) should only be present in the Camera-Ready Version of your paper. The version that you initially submit for review, ought to be double-blind. So, when initially submitting your paper, use:
%\author{Author information scrubbed for double-blind reviewing}
\author{Davide Ghia\inst{1} \and
Gabriele Ciravegna\inst{1,2} \and
Alkis Koudounas\inst{1} \and
Marco Fantini\inst{3,4} \and
Erika Crosetti\inst{4} \and
Giovanni Succo\inst{4,5} \and
Tania Cerquitelli\inst{1} }
% You may leave out the orcidID information, if you want to.
% Use \corr to indicate the corresponding author. Note the spacing around the \corr command. Only one author can be the corresponding author.

%N.B.: comment out the \authorrunning{} command for the double-blind version of your paper submitted for review. Later, if your paper is accepted, use the command for the Camera-Ready Version.
\authorrunning{Davide Ghia et al.}
% If there are two authors, write 'A.L. Benjamin and C.C. Broadus Jr.'
% If there are more than two authors, '[...] et al.' is used.

\institute{Politecnico di Torino, Turin, Italy 
\\\email{davide.ghia@studenti.polito.it} \and
{CENTAI Institute, Turin, Italy} \and
{San Feliciano Hospital, Rome, Italy} \and
{SCDU Otorinolaringoiatria, Head Neck Cancer Unit, Ospedale San Giovanni Bosco, Turin, Italy} \and
{Dipartimento di Oncologia, Università degli Studi di Torino, Turin, Italy}
%\email{\{gabriele.ciravegna,tania.cerquitelli\}@polito.it} 
}

\maketitle              % typeset the header of the contribution

\begin{abstract}
Voice disorders affect a significant portion of the population, and the ability to diagnose them using automated, non-invasive techniques would represent a substantial advancement in healthcare, improving the quality of life of patients. Recent studies have demonstrated that artificial intelligence models, particularly Deep Neural Networks (DNNs), can effectively address this task. However, due to their complexity, the decision-making process of such models often remain opaque, limiting their trustworthiness in clinical contexts.
This paper investigates an alternative approach based on Explainable AI (XAI), a field that aims to improve the interpretability of DNNs by providing different forms of explanations. Specifically, this works focuses on concept-based models such as Concept Bottleneck Model (CBM) and Concept Embedding Model (CEM) and how they can achieve performance comparable to traditional deep learning methods, while offering a more transparent and interpretable decision framework.

\end{abstract}
\section{Introduction}
\label{sec:intro}
In healthcare, the development of Artificial Intelligence (AI) offers new and more efficient approaches to various types of diagnosis. Non-invasive approaches represent an important example, significantly improving patients' quality of life. Although non-invasive AI-based approaches have been studied in different branches of medicine, such as skin cancer~\cite{esteva2017dermatologist}, diabetic retinopathy~\cite{gulshan2016development}, and atrial fibrillation~\cite{attia2019screening}, %voice analysis remains an under-explored domain. %cancer diagnosis \cite{huang2019lungcancer,ardila2019e2ecancer},  
their application to vocal disorder recognition still presents room for improvement. This task consists in the classification of pathological versus euphonic voices. Voice disorders represent an important pathology affecting a significant portion of the population, substantially impacting patient quality of life~\cite{bhattacharyya2014prevalence,cohen2010self,fantini2024rapidly,roy2005voice,spantideas2015voice}.

Very recently, Deep Neural Networks (DNNs) have started to demonstrate promising performances on this task \cite{Koudounas2024_Transf,koudounas2025mvp}, potentially suggesting that uncomfortable and invasive clinical procedures may be avoided through the simple analysis of voice recordings.
These complex models, mainly based on the Transformer architecture \cite{vaswani2017attention}, achieve competitive performances on the voice disorders detection task, thanks to the impressive correlation analysis capabilities of the attention module, which allows them to operate end-to-end on the raw voice data. Yet, their decision process remains predominantly opaque, and the model is perceived as a black box: its outputs are produced from the inputs without providing any form of explanation. Particularly in the context of healthcare, the lack of interpretability represents a crucial problem. Considering the sensitivity of the data involved, fully relying on the decisions of a non-transparent algorithm can raise both ethical and legal concerns \cite{palaniappan2024global}.

EXplainable AI (XAI) \cite{Mersha_2024XAI} aims to address the issue of interpretability in DNNs. Through various modalities and techniques, it aims to enhance the transparency and clarity of the decision process, using tools such as graphs, visual explanations, or saliency maps~\cite{pastor2024explaining}. 
Concept-based XAI (C-XAI) \cite{kim2018TCAV,Poeta2023CXAI_Review}, a promising and interesting subfield of Explainable AI, seeks to improve interpretability of the models by means of higher-level abstractions, a.k.a.  concepts. A \emph{concept} can be defined as a representation of the information learned by the model that is understandable by humans. 
Explaining model decisions in terms of such concepts can contribute to increasing user trust and acceptance of neural networks. 
Concept-based models \cite{Koh2020CBM} make a step further, directly providing the final prediction in terms of a combination of concepts. These models represent a significant improvement in terms of both transparency and interactivity, enabling human experts to intervene in concept representations to derive counterfactual predictions.

Unlike previous work, in this paper, we apply a concept-based approach to the voice disorder detection task. In order to collect annotations for the concepts, we used the Italian Pathological Voice (IPV) dataset introduced in \cite{Koudounas2024_Transf}, which contains the written anamnesis of each case. 
We then train two different types of concept-based models, a Concept Bottleneck Model (CBM) \cite{Koh2020CBM} and a Concept Embedding Model (CEM) \cite{Zarlenga2022CEM}, to learn both the concepts and the classification task. Through a comparison with a conventional end-to-end transformer-based approach, we demonstrate that concept-based models achieve competitive performance while significantly improving the interpretability of the decision process.
 
\section{Related Works}
\label{sec:formatting}
Voice disorders represent an important field of study within healthcare, as they affect a significant portion of the population, and are symptoms of serious illnesses, including both benign and malignant conditions, and neurodegenerative disorders~\cite{brunner2023prevalence,karabayir2020gradient,vieira2020voice}.
While invasive techniques have been demonstrated to be very effective for diagnosis \cite{Cohen2014LarStroboscopy1,Hamdan2023LarBiopsy,Pietruszewska2021LarStroboscopy2}, they often involve anesthesia, stressful exams, and risk of complications. A non-invasive approach known as Electroglottography (EGG) \cite{islam2022EGGexp,EGGutility}, which measures the contact between the vocal folds by applying electrodes to the patient's neck, minimizes the risks associated with invasive exams. 
Other techniques focus on the voice features extraction. Mel-Frequency Cepstral Coefficients (MFCCs) were used to train multi-layer perceptron (MLP) \cite{Salhi2008MLP}, while Convolutional Neural Networks (CNN) exploited MFCC cepstral spectrograms \cite{Peng2023VDCNN,CNN}. DNN, and in particular CNN, can also work directly on raw speech signal, as shown for 1D-CNNs. All these works demonstrated the feasibility of the task without, however, showing competitive performance with invasive approaches. 
On the contrary, transformer end-to-end architectures in \cite{ciravegna2024non,Koudounas2024_Transf,laquatra2025ger,ribas2023automatic} have started to show that voice analysis represents, with these architectures, a powerful non-invasive diagnostic approach. \cite{koudounas2025mvp,liu2025multimodal} showed that including different voice sources, such as sustained vowels and sentence readings,  further improves the detection performance. 

\vspace{2mm}
Still, all AI models work end-to-end on the raw data without providing almost any form of interpretability.
To address this challenge, various AI research groups have increasingly focused on explainable AI (XAI) \cite{Mersha_2024XAI}, and more recently, on C-XAI methodologies \cite{Poeta2023CXAI_Review}. Concept-based XAI techniques can be divided into two main categories: \emph{post-hoc} and \emph{explainable-by-design} approaches. Post-hoc methods operate on models trained conventionally and aim to detect the presence of concepts within their learned representations. In contrast, explainable-by-design methods integrate concepts directly into the architecture of the neural model, encouraging the learning of effective concept representations during training.
TCAV \cite{kim2018TCAV}, ACE \cite{ghorbani2019ACE}, and CaCE \cite{goyal2020CaCE} are examples of the former. These methods identify concepts within the learned representation of the model and quantify their influence on the final prediction. For the latter, existing approaches include supervised methods such as CBM \cite{Koh2020CBM}, CEM \cite{Zarlenga2022CEM}, and ProbCBM \cite{kim2023ProbCBM}; unsupervised methods such as SENN \cite{alvarezmelis2018SENN} and ProtoPNet \cite{Chen2019ProtoPNet}; and generative approaches such as Label-free CBM \cite{Oikarinen2023LabelFreeCBM} and LaBo \cite{Yang2023LaBo}. A comprehensive review of these techniques is provided in the work by \cite{Poeta2023CXAI_Review}.
Supervised explainable-by-design techniques proved to be the most effective in terms of improving the interpretability of the neural representation, while maintaining competitive performance. For instance, CBM and CEM models explicitly associate a node (CBM) or an embedding (CEM) to a specific concept. For these reasons, we adopted this approach in our work.
\section{Method}
\begin{figure*}[t]
  \centering
  \includegraphics[width=\textwidth]{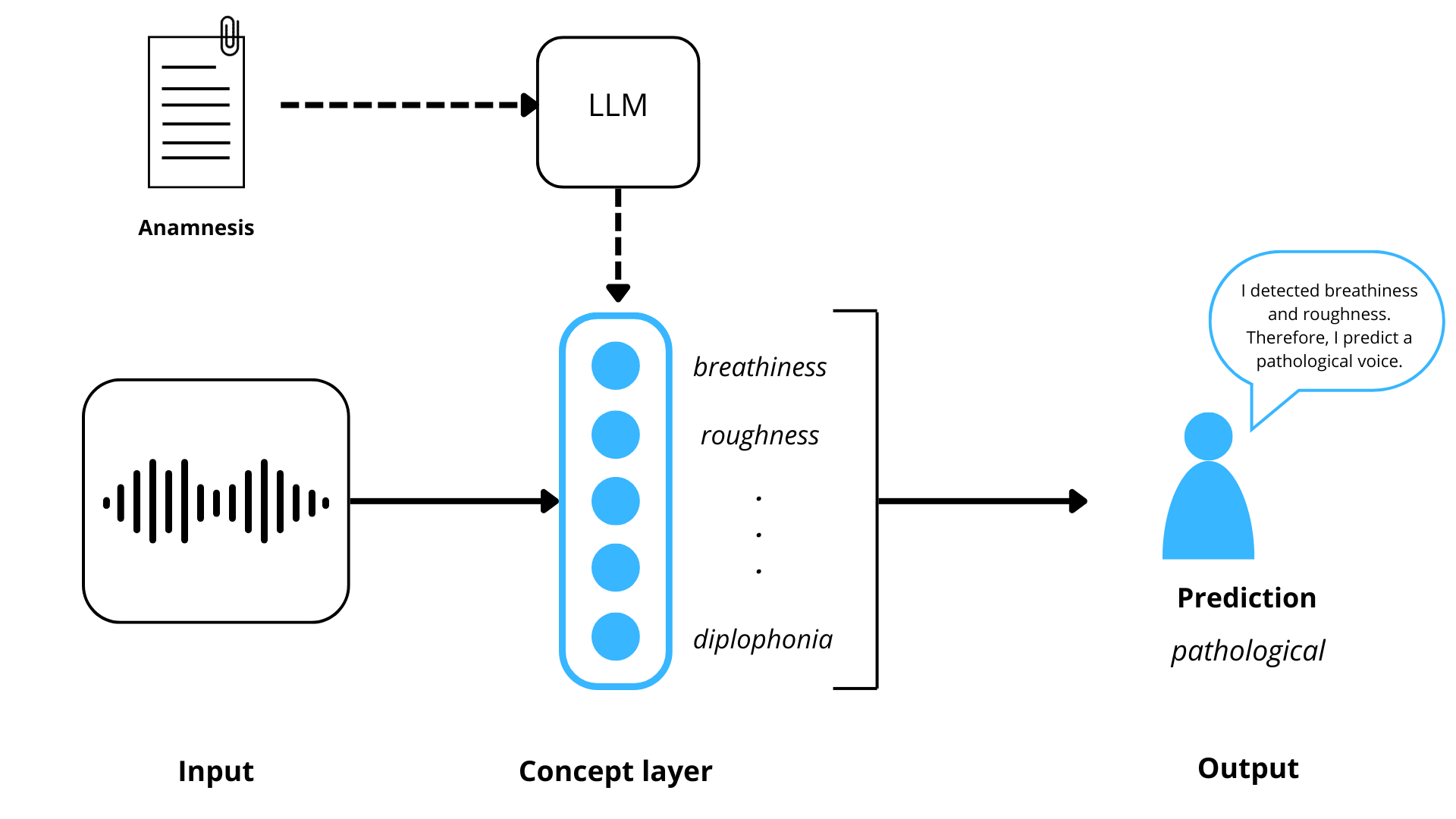}
  \caption{
    \textbf{Concept-based model framework}. The upper part of the figure shows how an LLM can provide concept annotation during training. We depict this process with dashed lines, as it is only required during training.  
    The bottom part represents how a model can learn to predict intermediate concepts to justify the final classification decision. 
  }
  \label{fig:conceptbasedframework}
\end{figure*}

This paper aims to create a concept-based model for detecting voice pathologies in an interpretable way. 
% The general intuition behind these types of models is illustrated in Fig. \ref{fig:conceptbasedframework}. 
The first requirement of supervised concept-based models is the availability of a dataset annotated with concepts.
Thus, the procedure followed in this paper consists in (i) extracting a set of annotated concepts from the textual files contained in the dataset with a Large Language Model (LLM) (see Section~\ref{sec:conc_ann}), and (ii) training a Concept Bottleneck Model and a Concept Embedding Model on the concepts extracted and feeding these concepts to a binary classifier for detecting a voice pathology (see Section~\ref{sec:cbm}). 
The encoder layers of the models are derived from a pre-trained audio transformer architecture (HuBERT \cite{hsu2021HuBERT}), while the classification head changes depending on the solutions adopted (e.g., CBM and CEM). 
Figure \ref{fig:conceptbasedframework} depicts the overall framework with the audio waveform and the anamnesis document serving as input data. During training, the first is used to extract the concept annotations by means of an LLM, while the second is given as input to an audio transformer to learn to output the concepts and the final predictions. At test time, the LLM and the anamnesis are not required anymore, since the audio model has now learned to predict the concepts. 
%A standard neural network, a pre-trained HuBERT as before, is also trained end-to-end with the raw waveform of the audio recordings to compare the results with the explainable models. 

%---------------------------------------------------------------------------

\subsection{Concept Annotation}
\label{sec:conc_ann}
The first step to train a concept-based model is to obtain a concept-annotated dataset. In our case, we utilize the IPV dataset~\cite{Koudounas2024_Transf}, which presents, for most cases, a PDF or Word file containing a written anamnesis of the patient. These files do not have a standardized structure, as they were not designed for concept prediction: they are written by medical experts in a textual form. Thus, after an initial technical consultation with experts and through a deep analysis of the data, both automated and manual, we identified a set of 14 candidate concepts that were reported in most anamneses\footnote{List of candidate concepts: \textit{smoking}, \textit{professional voice use}, \textit{dysphonia}, \textit{irregular mucosal wave}, \textit{mucous}, \textit{diplophonia}, \textit{strain}, \textit{roughness}, \textit{breathiness}, \textit{asthenicity}, \textit{phonastenia}, \textit{glottic hourglass configuration}, \textit{dysodia}, \textit{gender}.}.
By ``candidate concept'', we mean a concept that describes a particular characteristic of the patient, which may help distinguish between pathological and euphonic cases.

\begin{lstlisting}[language=java, frame=bt, caption=Example of prompt used to annotate concepts with the Gemini model. Following the anamnesis language\, the original prompt was in Italian\, but it has been translated here for the sake of readability. The four input/output examples used for few-shot learning are not reported for brevity., label=prompt]
Extract the following concepts from the provided text, returning a complete JSON object. Each concept must have exactly one of the allowed values.
[
	"smoking": allowed values: "yes"/"no", 
	"professional voice use": allowed values: "yes"/"no", 
	"dysphonia": allowed values: "no"/"light"/"light-moderate"/"moderate"/"severe", 
	"irregular mucous wave": allowed values: "yes"/"no",
	"mucous": allowed values: "pink"/"eutrophic"/"hyperemic"/"not present", 
	"diplophonia": allowed values: "yes"/"no", 
	"strain": allowed values: "yes"/"no", 
	"roughness": allowed values: "yes"/"no",
	"breathiness": allowed values: "yes"/"no",
	"asthenicity": allowed values: "yes"/"no",
	"phonasthenia": allowed values: "yes"/"no",
	"glottic hourglass configuration": allowed values: "yes"/"no",
	"dysodia": allowed values: "yes"/"no",
	"sex": allowed values: "male"/"female"
]
If the concepts are not explicitly mentioned in the text, their value must be "no" - or "not present" only when the anamnesis explicitly include "not present" among the allowed values (such as "mucous").

Important: Concepts like asthenicity, phonasthenia, roughness, or strain must be marked as "yes" only when they are explicitly mentioned in the text.

Here is an example of the output format to provide and strictly follow:
{
	"smoking": (your answer),
	"professional voice use": (your answer),
	"dysphonia": (your answer),
	"irregular mucous wave": (your answer),
	"mucous": (your answer),
	"diplophonia": (your answer),
	"strain": (your answer),
	"roughness": (your answer),
	"breathiness": (your answer),
	"asthenicity": (your answer),
	"phonasthenia": (your answer),
	"glottic hourglass configuration": (your answer),
	"dysodia": (your answer),
	"sex": (your answer)
}
# Example 1:
...
# Example 2:
...
# Example 3:
...
# Example 4:
...
\end{lstlisting} 

To extract the concepts from each anamnesis file, we used an LLM, which is capable of interpreting natural language. 
This approach was particularly suitable, as a method based on structured data was not applicable. 
We applied a technique known as \emph{few-shot prompting}, which involves including input-output examples within the prompt to better guide the model toward the desired output. In our case, we randomly selected four different files from the dataset as examples. We manually edited some of the original concept values to avoid repetition.
By using different combinations of concept values, we aimed to cover every possible value of each concept, in order to improve the model's ability to recognize them.
After partial editing, these examples were included in the prompt along with their corresponding manually annotated concepts. An example is shown in Listing~\ref{prompt}. 

We considered two distinct LLM models: \texttt{Mistral 7B-v0.1-hf} \cite{jiang2023mistral7b} and \texttt{Gemini-pro generative} \cite{geminiteam2025gemini}. Mistral is an open-source model that aims to find a compromise between performance and complexity, while Gemini is a powerful proprietary model developed by Google DeepMind, accessible only through API. In order to test the models' annotation capabilities and compare their results, we manually constructed a validation set: 69 files were selected from the original dataset, and each was humanly annotated with the corresponding concepts.

Since concept-based models such as CBM and CEM require binary concepts, we applied one-hot encoding to concepts with more than two possible values (e.g., \textit{dysphonia: absent, light, moderate, severe}). As a result, the total number of concepts in the candidate set increased from 14 to 20. After technical consultation with medical experts, we decided to exclude some concepts (e.g., \emph{mucous}) because they represented physical characteristics that could not be predicted through perceptual exams, but only via invasive procedures.
Other concepts, such as \emph{gender} or \emph{phonasthenia}, are information directly obtainable from the patient and do not require prediction. 
Finally, with the guidance of the medical experts, the concept \emph{dysphonia light-moderate} was merged with the concept \emph{dysphonia moderate} since it was an unnecessary distinction.
The final set (Table \ref{tab:concepts}) consists of 9 predictable concepts and 5 patient-provided concepts.  
The integration of these two groups is discussed in the following section.
Out of the 9 predictable concepts, 8 represent values within the GRBAS scale \cite{GRBAS}, a tool used by experts to evaluate the quality of voice. In this scale, ``G'' stands for the grade of dysphonia (represented by 4 concepts), ``R'' for roughness, ``A'' for asthenia, and ``S'' for strain.  

\begin{table}[t]
  \centering
  \caption{\label{tab:concepts}%
    Concept selection for the final concept set.
  }
  \begin{tabular}{lll}
    \toprule
    Predictable & Patient-provided & Excluded \\
    \midrule
    Dysphonia absent & Smoke &  Irregular mucous wave\\
    Dysphonia light & Professional use of voice & Mucous pink \\
    Dysphonia moderate & Gender & Mucous hyperemic\\
    Dysphonia severe &  Phonasthenia & Mucous eutrophic\\
    Diplophonia & Dysodia & Hourglass gottic configuration\\
    Strain & & \\
    Roughness & & \\
    Breathiness & & \\
    Asthenicity & & \\
    \bottomrule
  \end{tabular}
\end{table}

%---------------------------------------------------------------------------

\subsection{Concept-based architecture}
\label{sec:cbm}
\begin{figure*}[t]
  \centering
  \includegraphics[width=\textwidth]{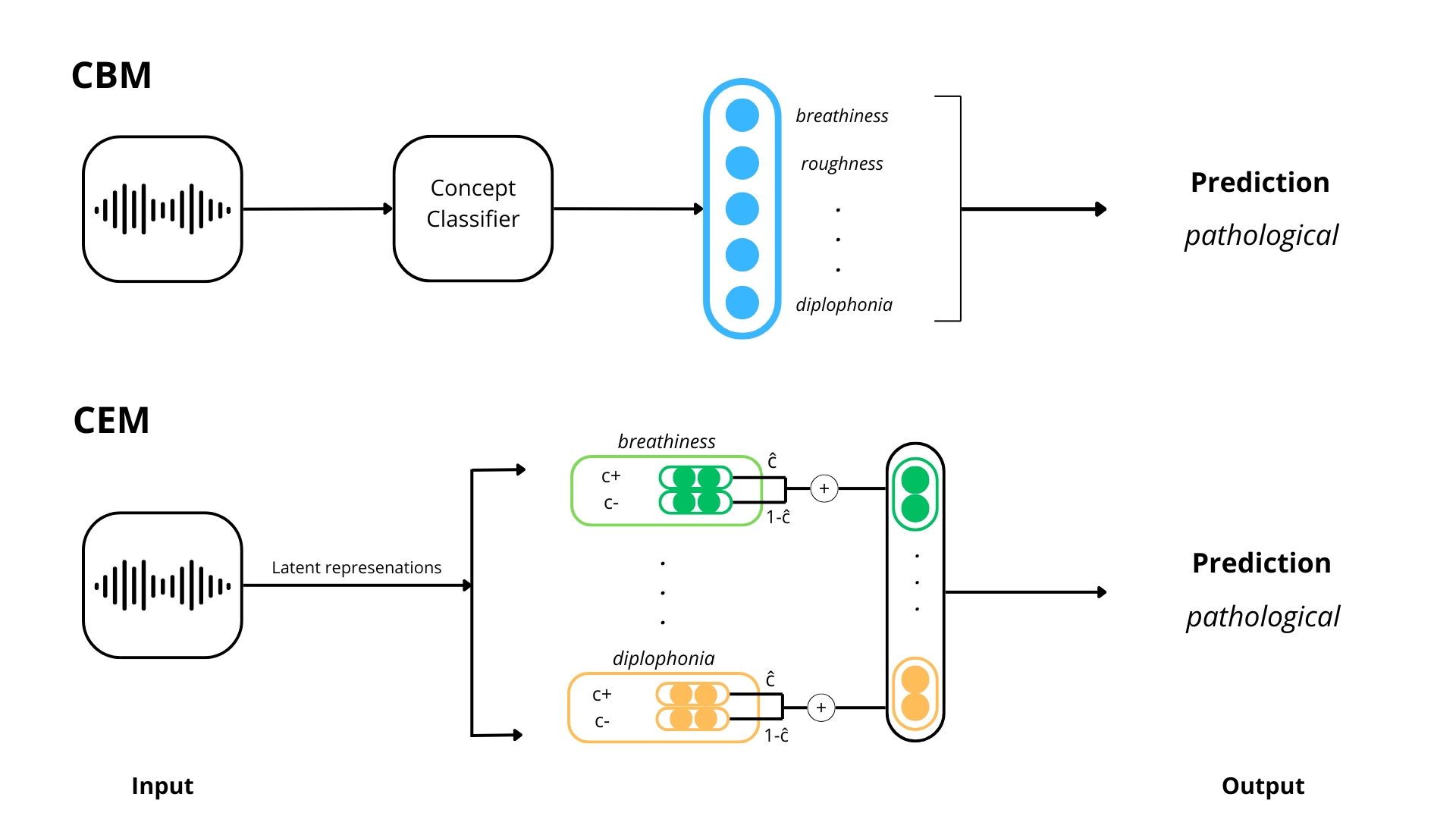}
  \caption{
    \textbf{CBM vs CEM architecture}. CBM predicts a logit for each concept in the bottleneck layer. In contrast, CEM represents each concept with two distinct embeddings: one for the active state and one for the inactive state; these embeddings are then weighted by the probability of the concept being active \(\hat{c}\) and inactive \(1-\hat{c}\), respectively.
  }
  \label{fig:cbmvscem}
\end{figure*}

In order to create an interpretable voice disorder detector, we propose a concept-based architecture, which can be adapted, with minor modifications, for both CBM and CEM approaches. An overview of the proposed scheme is reported in Figure \ref{fig:cbmvscem}.
To obtain meaningful audio embeddings, we used HuBERT \cite{hsu2021HuBERT}, an unsupervised pre-trained model specialized in speech representation, as a feature extractor. The use of a pre-trained model is particularly suitable given the limited size of our dataset. 

In CBM models, the function \(g(x) \), corresponding to the concept classifier, maps the input composed of $d$-dimensional space into the concept space, while \(f(c) \), the task classifier, maps the concepts into the class space. 
In our case, we adapt the concept classifier to work on the audio embedding $e = h(x)$, where $x \in X \subset \mathcal{R}^{dt}$ is the input data represented as a sequence of tokens of dimension $d$ and maximum length $t$, while the embeddings are represented as $e \in E \subseteq{R}^{m}$, where $m$ is the dimension of the embedding representing each sample. 
On the other side, we still require concepts and tasks to be valid for the entire sequence, hence \(c \in C \subseteq [0, 1]^k\) and \(y \in Y \subseteq [0,1] \). 

It is important to note that the task classifier bases its prediction solely on concepts and never has access to the audio signal or its representations.
In our CBM model, we add a classification head composed of a sequence of linear layers on top of the feature extractor. The final linear layer corresponds to the \emph{bottleneck layer}, where each node is associated with a unique concept. We obtain the bottleneck simply by resizing this layer to match the number of concepts to predict. In our case, the bottleneck layer outputs 9 different logits, corresponding to the 9 concepts to predict, on which a sigmoid function is applied. Through a threshold ($0.5$), the concepts are binary classified as present or absent.
The 9 predicted concepts \(\hat{c} = g(x)\) are concatenated with the 5 user-provided concepts $c"$ into a binary concept tensor \(\bar{c} = [\hat{c}, c"]\) that is fed to the task classifier \(\hat{y} = f(\bar{c}) \). 
A second classification head, also consisting of linear layers, takes the concept tensor as input and produces the final class prediction. The model is then trained as a \emph{joint bottleneck}, where the concept loss and the task loss are minimized at the same time using a combined loss:
\[
\hat{f},\, \hat{g} = \arg\min_{f, g} \sum_i \left[ \mathcal{L}_Y(f(g(x^{(i)})), y^{(i)}) + \sum_j \lambda \, \mathcal{L}_{C_j}(\hat{c}; c^{(i)}) \right]
\]
where \( \mathcal{L}_Y \) represents the task loss, while \( \mathcal{L}_{C_j} \) corresponds to the concept loss. The hyperparameter \( \lambda \) regulates the weight of the concept loss in the total loss computation.

The structure of the CEM model is nearly identical to that of the CBM model. The bottleneck layer is replaced by an \emph{embedding layer}, hence $\mathbf{c} = g_{CEM}(x)$, where $\mathbf c \in \mathbf{C} \subset \mathcal{R}^{kh}$, where $h$ is the dimension of each concept embedding\footnote{CEM does also provide concept prediction $\hat{c}$ and condition the concept embedding through convex combination of the concept prediction with the positive and negative concept embeddings \(\mathbf{c} = g(x) = \hat{c}\cdot \mathbf{c}^+ + (1-\hat{c})\cdot \mathbf{c}^- \).}. 
As shown in Fig. \ref{fig:cbmvscem}, instead of a direct one-to-one association between nodes and concepts, in this layer each concept is represented by two distinct embeddings, one corresponding to its active state and the other to its inactive state. The embedding layer also predicts the probability of each concept being active, over the concatenation of the positive and negative embeddings $\hat{c}_i = s[(\mathbf{c}^+_i, \mathbf{c}^-_i])$, which are predicted by means of a learned scoring function shared across concepts. The final classification head receives as input the concatenation of the concept embeddings $f(\mathbf{c})$, which are mapped into a shared latent space.

\section{Experiments}
We design our experiments to evaluate whether concept-based models (CBM, CEM) can match the performance of end-to-end models while offering greater interpretability in pathological voice classification from clinical recordings.

\subsection{Dataset}
The dataset used in the experiments is the Italian Pathological Voice (IPV). It consists of recordings, anamnesis, and exam results of 513 patients, collected from several private phoniatric and speech-therapy practices and hospitals in Italy. The dataset presents 170 euphonic cases and 340 dysphonic cases. For each patient, the dataset contains the recording of the reading of five phonetically balanced sentences derived from the Italian adaptation of the CAPE-V \cite{capev}, and the recording of the sustained production of the vowel /a/. In addition to the AVQI exam results, which are not used in this paper, there are also the files containing the anamnesis of the patient, written directly by the doctor who performed the visit. Out of 513 cases, only 385 present a written anamnesis, out of which 134 are euphonic cases and 251 pathological cases. Since the dataset size is very limited, especially for large models as transformers, we performed 10-fold cross-validation to ensure robustness.

%---------------------------------------------------------------------------

\subsection{Experimental Details}

\textbf{Concept annotation.} We tested two distinct LLMs, the open-source \texttt{Mistral 7B-v0.1-hf} and the closed-source \texttt{Gemini-pro generative} models on the manually annotated test set. The experiments were conducted on the annotations of the original concept candidate set, composed of the 14 concepts, in order to test the capability of the LLMs to recognize all the selected concepts, both predictable and patient-provided. Given that a structured and coherent output was needed, the temperature of both models was set to 0.1. To better evaluate the models, we used both concept accuracy and the macro average F1 score. To facilitate the comparison, we also considered the total number of annotation errors.

\vspace{2mm}
\noindent \textbf{Voice pathologies detection}. The voice pathologies classification experiments were conducted only on cases that included a written anamnesis. We chose to use the recordings of the CAPE-V sentences, as this type of data contains more significant information. %Nevertheless, the most significant experiments were also repeated on recordings of sustained vowels to enable comparison. 
Each recording was padded to the maximum length of the dataset, as HuBERT requires inputs of equal length. All audio signals are resampled at 16 kHz. HuBERT model outputs a representation for each frame\footnote{A frame is a fixed-length time interval over the audio signal, from which the model extracts a representation. Typically, the length of this interval is 20ms.}. To obtain a unique representation, we applied max pooling over the model output. This technique was preferred to other methods, such as average pooling, as vocal characteristics tend to manifest as local peaks.
In order to have a baseline comparison for the concept-base models, we also trained a conventional neural network, i.e., a HuBERT architecture with a classification head composed of sequential linear layers, with layer dimensions progressively decreasing: 1-layer [128], 2-layer [256,128], 3-layer [512,256,128].
We set the learning rate for HuBERT to 5e-5, while for the concept-based models, we used 5e-5 for the concept classifier and 5e-4 for the task classifier.
For the CEM model, we employed an embedding size $h=16$. We trained the models for 30 epochs, with early stopping criteria.
We also applied a warm-up strategy: we froze the task loss for the first 2 epochs of training, leaving active only the concept loss. By doing so, the model begins to learn the task after learning better concept representations.

We evaluate the models using concept accuracy, task accuracy, and task F1 score. Concept accuracy is crucial to assess whether the model correctly learned effective representations of concepts. To prioritize this aspect during training, we weighted the total loss as 0.9 ($\lambda$) for the concept loss and 0.1 for the task loss. The F1 score is important to evaluate if the model learned to distinguish between both classes, especially given the significant class imbalance in the original dataset.

%---------------------------------------------------------------------------

\section{Results}
We first report the effectiveness of the concept annotations in order to ensure that the concept-based models would train on reliable data. We then analyze the performance of the CBM model and CEM model, comparing the results with those obtained by training a standard transformer model. 

\vspace{2mm}
\noindent \textbf{Concept annotation.} As shown in Table \ref{tab:annotations}, both LLMs achieve satisfying results. It is important to note that some errors occur in concepts that were either vaguely described in the clinical reports or whose annotation heavily relied on interpretation. For instance, if the degree of dysphonia is not explicitly stated, the corresponding concept value must be inferred.
\begin{table}[t]
  \centering
  \caption{\label{tab:annotations}%
    \textbf{Concept annotation}. Results of the concept annotation performed by the two considered LLMs.
  }
  \begin{tabular}{lccc}
    \toprule
    Model & Concept Acc & Concept F1  macro & Total Errors \\
    \midrule
    Mistral 7B & 0.9706 & 0.9670 & 42\\
    Gemini-pro & \textbf{0.9865} & \textbf{0.9810} & \textbf{20} \\
    \bottomrule
  \end{tabular}
\end{table}
The Gemini model performs slightly better than Mistral in terms of accuracy and F1 score. The difference in performance increases if we consider the errors on the total number of concept predictions. Based on these considerations, Gemini was selected to perform the full annotation of the original dataset.

% \paragraph{Baseline} 
%\begin{table}[t]
%  \centering
%  \caption{\label{tab:classifiers}%
%    \textbf{Baseline}. Comparison between the results of different configuration of the conventional neural network. The F1 score for each class is reported in parentheses. 
%  }
%  \begin{tabular}{lccc}
%    \toprule
%    \textbf{Classifier} & \textbf{Val Accuracy} & \textbf{Test Accuracy} & \textbf{Test F1 score} \\
%    \midrule
%    1-LAYER conf & 80.36\% \textnormal{\scriptsize(±.)} & 82.46\% \textnormal{\scriptsize(±.)} & \textbf{0.76} (0.64-0.88) \\
%    2-LAYER conf & 86.31\% \textnormal{\scriptsize(±.)} & \textbf{84.80\%} \textnormal{\scriptsize(±1.00)} & \textbf{0.76} (0.62-0.90) \\
%    3-LAYER conf & \textbf{86.91\%} \textnormal{\scriptsize(±.)} & 81.87\% \textnormal{\scriptsize(±.)} & 0.70 (0.50-0.89)\\
%    \bottomrule
%  \end{tabular}
%\end{table}
%As we can see in Table \ref{tab:classifiers}, 
%The 3-layer structure over-parametrize the output of HuBERT (last hidden layer dimension = 768). 

\begin{table}
  \centering
  \caption{\label{tab:compare}%
    \textbf{Concept-based models}. Results of the concept-based model, compared to the conventional DNN and the Ideal CBM.
  }
  \begin{tabular}{lccc}
    \toprule
    \bfseries Model & \bfseries Concept Accuracy & \bfseries Task Accuracy & \bfseries Task F1 \\
    \midrule
    HuBERT & - & \textbf{0.9133} \textnormal{\scriptsize(±0.0414)} & \textbf{0.9085} \textnormal{\scriptsize(±0.0476)}\\
    CBM & 0.8443 \textnormal{\scriptsize(±0.0346)} & 0.8776 \textnormal{\scriptsize(±0.0247)} & 0.8566 \textnormal{\scriptsize(±0.0175)}\\
    CEM & \textbf{0.8450} \textnormal{\scriptsize(±0.0205)} & 0.8730 \textnormal{\scriptsize(±0.0385)} & 0.8599 \textnormal{\scriptsize(±0.0168)}\\
    \midrule
    Ideal CBM & - & 0.9052 \scriptsize(±0.0259) & 0.8953 \scriptsize(±0.0309)\\
    \bottomrule
  \end{tabular}
\end{table}

\vspace{2mm}
\noindent 
\textbf{Voice pathologies prediction.} 
As end-to-end baseline, the best-performing architecture overall is the 2-layer classification head, which achieves a Task Accuracy of 0.9133 and an F1 score of 0.9085, as reported in Table \ref{tab:compare}. We adopt this model as the baseline for the comparison with the concept-based models. 

We trained the concept-based models using the annotations produced by the Gemini model. The concept accuracy was calculated on the 9 predicted concepts. To ensure that the identified concepts are effective in distinguishing the two classes, we trained a model (referred to as Ideal CBM), which receives the true concept values as input and predicts the corresponding class.
In Table \ref{tab:compare} we report the results of CBM and CEM models, the best-performing standard HuBERT model trained end-to-end for the voice pathologies detection task, and the Ideal CBM. 
It is worth noting that, with an ideal concept accuracy of 100.00\% (IdealCBM), the performance of the CBM is very close to that of HuBERT, highlighting the effectiveness of the selected concepts for the classification task.

The actual concept accuracy achieved by CBM and CEM is approximately 84.50\%. This can be attributed to the ambiguity of some annotations, which can cause erroneous predictions.
Concept-based models achieve results comparable to those of the end-to-end approach in terms of both accuracy and F1 score, maintaining high classification effectiveness and introducing a significant level of interpretability. 
A pathological voice can be justified with a specific degree of dysphonia and by the presence of features such as \textit{breathiness}, \textit{roughness}, \textit{diplophonia}, and \textit{strain}. 
CBM slightly outperforms CEM in task accuracy despite having lower concept accuracy. Overall, however, the performance of the two models remains nearly identical.
\section{Conclusion}
In this paper, we proposed a concept-based approach to the voice disorders prediction task, in order to reach a high degree of interpretability, crucial in a delicate field as healthcare. 
We demonstrated the effectiveness of these models in recognizing both vocal concepts and pathological voices, achieving performance comparable to the conventional approach.

\paragraph{Future work} A possible direction going forward could be considering not only the read sentences but also sustained vowels as input data, utilizing transformer models pre-trained for this kind of data~\cite{voc2vec,arch}. 
An interesting direction for future development could also be the enhancement of the explanation interface. Concept predictions could be linked to a generative model (e.g., LLM) to produce a report-like document, including a description of the vocal analysis, suggested guidelines, and other clinically relevant information. 
Furthermore, the concept-based models employed in this paper could be evaluated on different datasets, such as AVFAD \cite{Jesus17AVFAD}. Concept annotations are required only during training, making this approach potentially applicable to different voice disorder datasets.

\begin{credits}
%\subsubsection{\ackname} %A bold run-in heading in small font size at the end of the paper is used for general acknowledgments, for example: This study was funded by X (grant number Y).

\subsubsection{\discintname}
Authors have no competing interest.
%It is now necessary to declare any competing interests or to specifically state that the authors have no competing interests. Please place the statement with a bold run-in heading in small font size beneath the (optional) acknowledgments, for example: The authors have no competing interests to declare that are relevant to the content of this article. Or: Author A has received research grants from Company W. Author B has received a speaker honorarium from Company X and owns stock in Company Y. Author C is a member of committee Z.
\end{credits}
\end{document}